\documentclass[10pt,conference]{IEEEtran}
\usepackage{graphicx}
\usepackage{booktabs}
\usepackage{subcaption}
\usepackage{amsmath, amsfonts, amssymb}
\usepackage{hyperref}
\usepackage{braket}
\usepackage{listings}
\usepackage{xcolor}
\usepackage{physics}
\usepackage{algorithm}
\usepackage{algpseudocode}
\algrenewcommand\algorithmicindent{0.4em}
\usepackage[most]{tcolorbox}
\tcbuselibrary{theorems, skins}

\definecolor{blue}{HTML}{435BEC}
\definecolor{peach}{HTML}{F07362}
\definecolor{purple}{HTML}{6A4C93}
\definecolor{coral}{HTML}{FF6F61}
\definecolor{gray}{HTML}{808080}

\tcbset{
  boxedthm/.style={
    enhanced,
    breakable,
    sharp corners,
    boxrule=0.6pt,
    left=1pt, right=1pt, top=1pt, bottom=1pt,
    fonttitle=\bfseries,
    before skip=3pt,
    after skip=3pt,
    separator sign={.},
  }
}

\newtcbtheorem[number within=section]{prediction}{Prediction}{
  boxedthm,
  colback=gray!8,
  colframe=gray,
}{pred}

\title{Pre-Asymptotic Trainability in Photonic Variational Circuits under Postselection}

\author{
  \IEEEauthorblockN{Yichen Xie}
  \IEEEauthorblockA{\textit{La Salle College}\\
    Hong Kong\\
    s21325@lsc.hk}

  \and

  \IEEEauthorblockN{Cassandre Notton}
  \IEEEauthorblockA{\textit{Quandela Quantique Inc.}\\
    Canada\\
    cassandre.notton@quandela.com}

  \and

  \IEEEauthorblockN{Jean Senellart}
  \IEEEauthorblockA{\textit{Quandela}\\
    France\\
    jean.senellart@quandela.com}
}

\begin{document}

\maketitle

\begin{abstract}
Barren plateaus (BP) in variational quantum circuits are commonly attributed to strong mixing dynamics that cause gradient variance to vanish exponentially with system size. Passive photonic circuits challenge this picture. Because observables and postselection maps generically extend beyond the algebra's reachable image, we show that trainability is governed by how postselection redistributes weight across the full irreducible decomposition of the operator space. Building on recent representation-theoretic second-moment techniques for passive linear optics, we express postselection through the pulled-back observable and obtain a closed-form Haar-averaged gradient variance, exact for fixed observables. The resulting scaling predictions are tested by statevector simulation of the nonlinear Bhattacharyya loss across allow-bunching, collision-free, and dual-rail postselection, using a four-test protocol (AICc, sliding-window, crossover, and truncation analyses) to separate polynomial from exponential scaling in finite data. Allow-bunching and collision-free filtering remain polynomial-consistent across all tested sizes and initializations, while dual-rail postselection induces a genuine BP at a rate roughly an order of magnitude below the 2-design value. 

\end{abstract}



\section{Introduction and Related Work}
\label{sec:intro}

Variational quantum algorithms (VQAs) pair parameterized quantum circuits with classical optimization loops and are widely studied as candidates for near-term quantum utility in chemistry, optimization, and machine learning ~\cite{Peruzzo2014,Cerezo2021VQA,Bharti2022}. Their scalability, however, is fundamentally constrained by the barren plateau (BP) phenomenon: gradient variance vanishes exponentially with system size for broad circuit and cost-function classes, making gradient-based training intractable ~\cite{McClean2018}. BPs have since been linked to circuit expressibility~\cite{Holmes2022}, global cost functions~\cite{Cerezo2021cost}, hardware noise~\cite{Wang2021NIBP}, and excessive entanglement in the input state~\cite{Larocca_2025,cunningham2025investigatingmitigatingbarrenplateaus}. Moreover, several of the proposed remedies appear to entail classical simulability, calling the very premise of quantum advantage into question~\cite{Cerezo2025simulability}.

A Lie-algebraic theory~\cite{Fontana2024adjoint} has since unified the known BP sources-circuit expressibility~\cite{Holmes2022}, global cost functions~\cite{Cerezo2021cost}, noise~\cite{Wang2021NIBP} (unified via \cite{Ragone2024}'s treatment of algebraic decoherence), and input-state entanglement~\cite{Larocca_2025,cunningham2025investigatingmitigatingbarrenplateaus} under a single variance formula. The variance of the loss is governed by the dimension of the dynamical Lie algebra (DLA) generated by the circuit's parameterized gates, together with algebraic purities of the initial state and observable.

Photonic quantum computing is a mature near-term platform: VQAs have been demonstrated on universal reconfigurable hardware~\cite{Maring2024,Baldazzi2025} and even deployed on industrially relevant PDE problems, where BP is already being engineered. For instance, a 2D fracture mechanics VQA was recently ran on Quandela's Ascella processor, bypassing BP up to 19 qubits via a heuristic cascaded remeshing warm-start~\cite{remond2025quantum}. Yet BP in photonic architectures remains largely unexplored. In this work, we focus on \emph{passive} linear optical circuits, where programmable MZI meshes~\cite{Reck1994,Clements2016,Marchisio2025} implement a mode-unitary $S \in U(m)$ on $m$ modes. On the $n$-photon Fock sector $\mathcal{H}_{m,n}$, $S$ induces $\pi_{m,n}(S)$ with DLA $\mathfrak{u}(m)$ of dimension $m^2$, while the Hilbert space dimension $\binom{m+n-1}{n}$ grows exponentially. This algebraic gap suggests photonic trainability is regime-dependent rather than uniformly obstructed. A further departure from gate-model intuition \cite{McClean2018} is that depth is not meaningful here, since products of mode-unitaries are themselves mode-unitaries and stacking MZI layers is algebraically redundant. The meshes we use are instead universal~\cite{Reck1994,Clements2016} within $U(m)$, distinct from the computational universality of KLM-style protocols, which additionally require encoded logical states, ancillas, measurement-induced nonlinearities, and feed-forward.

The remaining architectural axis is postselection, used to enforce encodings such as dual-rail qubits~\cite{Kok2007,Knill2001,Maring2024}, filter collision-free outcomes, or implement measurement-induced gates~\cite{Knill2002}. A postselection map $K$ pulls back the observable to $O^{(K)} = K^\dagger O K$, reshaping its projection onto the reachable Lie-algebraic operator modules and hence the variance, making postselection the natural axis along which photonic trainability varies. 

Prior work on bosonic trainability has focused on continuous-variable architectures, where ~\cite{Zhang_2024} identified an energy-dependent BP: gradient variance decays exponentially in mode count but only polynomially in per-mode energy. ~\cite{Monbroussou_2025} studied Hamming-weight preserving circuits and showed that trainability scales with the dimension of the preserved subspace. Using related representation-theoretic tools, ~\cite{mhiri2026boson} established anti-concentration of boson-sampling output distributions beyond the dilute regime, which is a distinct notion of concentration from the gradient variance considered here. However, a systematic study of how postselection regimes in discrete-variable photonic architectures affect gradient concentration has been lacking.

\textbf{Contributions.} We extend the representation-theoretic second-moment framework of \cite{mhiri2026boson,arienzo2025bosonic} to postselected photonic circuits, expressing postselection through the pulled-back observable $O^{(K)} = K^\dagger O K$ and deriving the corresponding gradient variance for fixed observables. We compare gradient-variance statistics across 3 postselection regimes and 3 initialization strategies, using a model-discrimination protocol built to separate polynomial from exponential scaling from finite data. We find dual-rail postselection induces exponential concentration that is invisible on the commonly accessible range ($N \le 10$) but robust once extended to $N = 24$, while collision-free filtering stays polynomial. We trace the divergence to how each postselection map redistributes weight across irreducible modules rather than to subspace size.

\section{Methodology}
\subsection{Mathematical Framework}
\label{sec:math-fram}
We now specialize the Lie-algebraic trainability framework to passive photonic circuits with postselection. Using the notation of Section~\ref{sec:intro}, $n$ indistinguishable photons in $m$ optical modes evolve under a passive linear circuit $S(\theta)\in U(m)$, lifted to the Fock sector $\mathcal{H}_{m,n}$ via $\pi_{m,n}$, with DLA $\mathfrak{g}=\mathfrak{u}(m)=\mathfrak{su}(m)\oplus\mathfrak{u}(1)$.

\subsubsection{Variance Formula}

For a parameterized circuit $g=\pi_{m,n}(S(\theta))$, input state $\rho$, and observable $O$, the cost function is $f(g)=\mathrm{Tr}[\pi(g)\rho\,\pi(g)^\dagger O]$. Postselection via a Kraus map $K$ yields
\begin{equation}
  f(g) = \mathrm{Tr}\!\left[\pi(g)\rho\,\pi(g)^\dagger\, O^{(K)}\right],
  \quad O^{(K)} := K^\dagger O K.
  \label{eq:postselected_obj}
\end{equation}

\subsubsection{Irreducible Decomposition of the Photonic Operator Space}

We specialize the variance analysis to passive photonic circuits by exploiting the representation-theoretic structure of the $n$-photon operator space. \cite{arienzo2025bosonic} establishes the complete reducibility of passive linear optics adjoint representation (Lemma 1). They show that, under the adjoint action $\omega_n = \varphi_n \otimes \varphi_n^{*}$ of a passive unitary $U(m)$ lifted via $\pi_{m,n}$, the space of operators acting on the $n$-photon Fock sector, $W_n = \mathcal{H}_{m,n}\otimes\mathcal{H}_{m,n}^{*}$, decomposes into $n+1$ inequivalent irreducible $U(m)$-modules~\cite{mhiri2026boson}:

\begin{equation}
  W_n \;\simeq\; \bigoplus_{k=0}^{n} \lambda_k^{(n)},
  \qquad
  d_k^{(n)} \;=\; \frac{2k+m-1}{m-1}\binom{k+m-2}{k}^{\!2},
  \label{eq:irrep_decomp}
\end{equation}
where $\lambda_0^{(n)}$ is the trivial module and $d_k^{(n)}$ denotes the dimension of the irreducible component $\lambda_k^{(n)}$. Each $\lambda_k^{(n)}$ is generated from a primitive lowest-weight subspace by repeated application of a raising map $R$ acting between adjacent photon-number sectors, with an associated lowering map $L$; $\{L,R\}$ generate an $\mathfrak{sl}_2(\mathbb{C})$ action on the graded operator space that commutes with the $U(m)$ action (Howe-duality Proposition 1 in \cite{mhiri2026boson}), yielding an explicit recursive formula for the projector $P_k^{(n)}$ onto each $\lambda_k^{(n)}$, which we realize concretely on the Fock-sector basis in Appendix~\ref{app:gpurity_computation}.

\subsubsection{Exact Second Moment Under Haar-Random Interferometers}

Averaging the postselected cost function over Haar-random interferometers $U\sim U(m)$ and exploiting the decomposition in Eq.~\eqref{eq:irrep_decomp}, Schur's lemma \cite{bartlett2007reference} gives the exact second moment
\begin{equation}
  \mathbb{E}_{U\sim U(m)}\!\big[f_U(\rho,O)^2\big]
  \;=\;
  \sum_{k=0}^{n}
  \frac{\big\|P_k^{(n)}(\rho)\big\|_2^2\;
        \big\|P_k^{(n)}\!\big(O^{(K)}\big)\big\|_2^2}
       {d_k^{(n)}},
  \label{eq:second_moment}
\end{equation}
where the $k=0$ term reproduces the squared mean $\mathbb{E}_U[f_U(\rho,O)]^2$ (Proposition 2 in \cite{mhiri2026boson}). The variance of the postselected cost is therefore
\begin{equation}
  \mathrm{Var}_{U}[f]
  \;=\;
  \sum_{k=1}^{n}
  \frac{\big\|P_k^{(n)}(\rho)\big\|_2^2\;
        \big\|P_k^{(n)}\!\big(O^{(K)}\big)\big\|_2^2}
       {d_k^{(n)}}.
  \label{eq:photonic_var}
\end{equation}

Equation~\eqref{eq:photonic_var} is the variance of the cost. For the gradient, writing
$U=U_2e^{i\theta_\mu G_\mu}U_1$ with $G_\mu\in\mathfrak{u}(m)$, the same
mean-vanishing argument gives $\mathbb{E}[\partial_\mu f]=0$, and averaging
$G_\mu$ over an orthonormal basis of $\mathfrak{u}(m)$ replaces the generator
by the Casimir of the adjoint action $\omega_n$, which by Schur's lemma acts on
$\lambda_k^{(n)}$ as the scalar $c_k = 2k(k+m-1)$.\footnote{$c_k$ is the $\mathfrak{su}(m)$ Casimir eigenvalue on $\lambda_k^{(n)}$, obtained from the standard $\mathfrak{gl}(m)$ formula by subtracting the $\mathfrak{u}(1)$ contribution $(\sum_i\lambda_i)^2/m$.} This gives the exact generator-averaged gradient variance
\begin{equation}
  \overline{\mathrm{Var}[\partial_\mu f]}
  = \frac{1}{m^2}\sum_{k=1}^{n}
    \frac{c_k\,\|P_k^{(n)}(\rho)\|_2^2\,
          \|P_k^{(n)}(O^{(K)})\|_2^2}{d_k^{(n)}},
  \label{eq:var_grad_casimir}
\end{equation}
Equation~\eqref{eq:var_grad_casimir} differs from Eq.~\eqref{eq:photonic_var} only by the module weights $c_k/m^2$. Since $c_k/m^2 = 2k(k+m-1)/m^2$ is increasing in $k \in [1,n]$, these weights lie between $2/m$ and $2n(n+m-1)/m^2$; for all configurations considered here ($n\le m/2$) the upper bound is below $3/2$. As both variances are sums of non-negative module contributions, their ratio is bounded by these extremes, so cost and gradient variance differ by at most a factor linear in $m$: they share the same exponential decay rate and differ only in polynomial prefactors. We therefore report the exactly computable cost variance of Eq.~\eqref{eq:photonic_var} as a faithful diagnostic for distinguishing exponential from polynomial scaling. 

\begin{prediction}{No-Code Polynomial Trend}{nocode}
\label{pred:nocode}
Without code postselection,
$O^{(K)}$ retains substantial weight in low-$k$ modules, whose dimensions
$d_k^{(n)} = \Theta(m^{4k})$ are polynomial in $m$ at fixed $k$. We therefore predict that at fixed photon number and increasing mode count the gradient variance follows $\widehat{V}(m)\in O(m^{-c})$ rather than  suppression, in contrast to Prediction II.2..

\end{prediction}

\begin{prediction}{Code-Postselection Exponential Trend}{code}
\label{pred:code}
Under $r$-rail code postselection with $N$ logical units, we expect the weight of $O^{(K)}$ to concentrate on modules of index $k$ scaling with $N$, whose dimension $d_k^{(n)}$ grows exponentially in $N$. We therefore expect the gradient variance to become compatible with an exponential trend, of the form $\exp(-aN)$ for some $a>0$, once the logical system size is large enough. In that regime, the behavior would be consistent with a BP in the number of logical units.
\end{prediction}

\begin{figure}[h]
    \centering
    \includegraphics[width=0.98\linewidth]{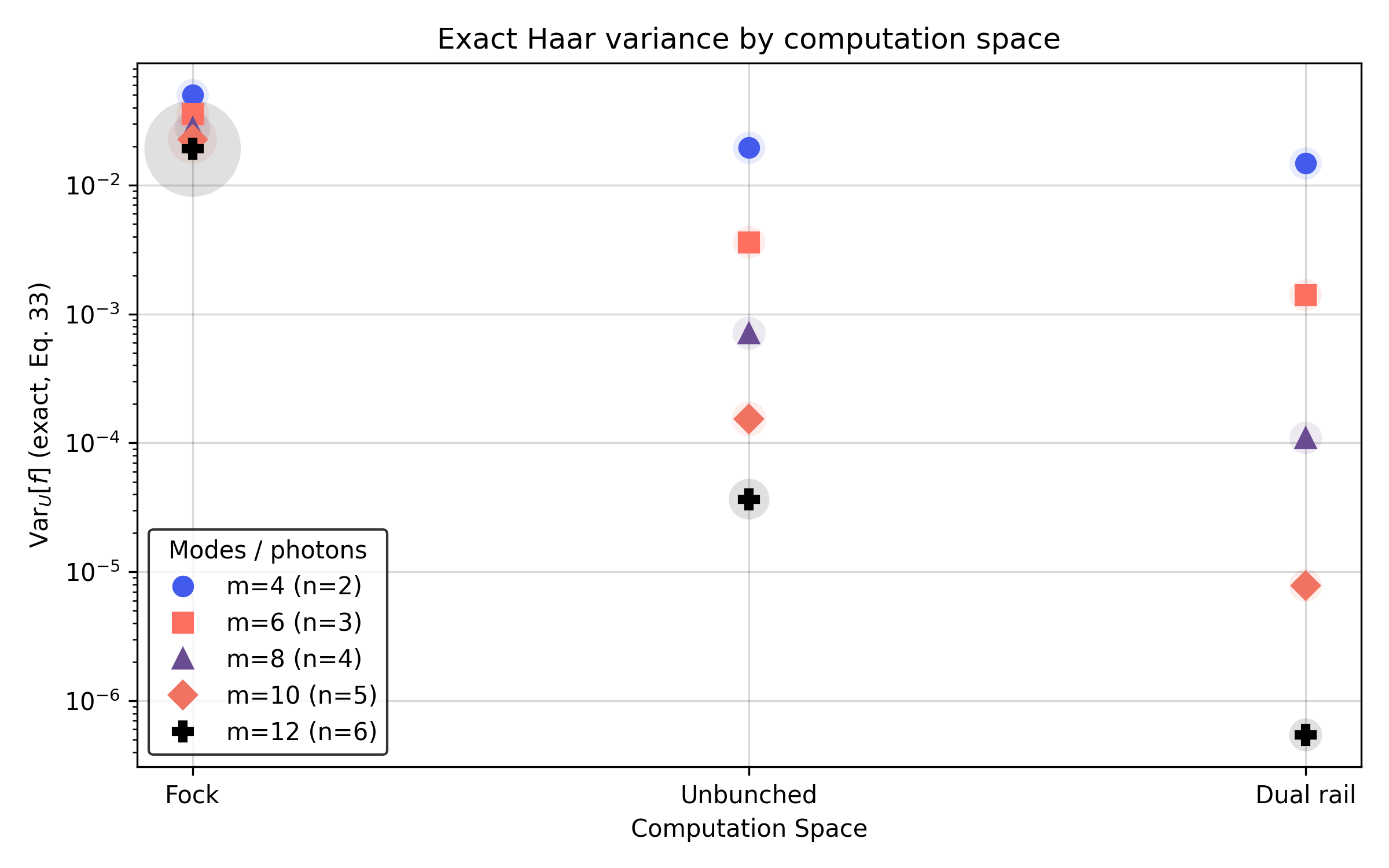}
    \caption{Exact cost variance $\mathrm{Var}_U[f]$ (Eq. (33) in \cite{mhiri2026boson}) as a function of computation space. Stricter postselection systematically decreases the variance, consistent with stronger variance suppression under rail-code projection. Marker size is proportional to subspace dimension.}
    \label{fig:var_exact}
\end{figure}

Figure~\ref{fig:var_exact} shows the exact cost variance of Eq.~\eqref{eq:photonic_var} as a function of computation space for $m\in\{4,6,8,10,12\}$ with $n=m/2$ photons. The hierarchy $\mathrm{Var}^{\,\textsc{fock}} > \mathrm{Var}^{\,\textsc{unbunched}} > \mathrm{Var}^{\,\textsc{dual\_rail}}$ is observed across the tested mode counts, with the separation widening with $m$, consistent with a systematic geometric effect of postselection.

\subsection{Experimental Setup}

\subsubsection{Circuit Family}
All photonic experiments use rectangular MZI meshes~\cite{Clements2016} capable of representing arbitrary passive mode-unitaries $S \in U(m)$ on $m$ modes,
parameterized by beam splitter angles and phase shifts. This is universality over passive interferometers, not universality for fault-tolerant or measurement-based quantum computation. The input state is a fixed Fock state with one photon per designated input mode, consistent with single-photon source architectures~\cite{Maring2024,Baldazzi2025}. All photonic runs use a single mesh layer, $\mathrm{depth}=1$ as the depth is not an independent expressivity axis for this circuit family (Section~\ref{sec:intro}).

\subsubsection{Postselection Regimes}
\label{sec:postselection_regimes}
We consider three postselection regimes. In the \emph{allow-bunching}, $K = \mathbb{I}$ and the full $n$-photon sector is retained. In the \emph{collision-free} setting, $K = K_{\mathrm{cf}}$ projects onto:
\begin{equation}
\Omega_{\mathrm{cf}} = \{s \in \mathbb{Z}_{\geq 0}^m : s_j \in \{0,1\},\, \textstyle\sum_j s_j = n\},
\label{eq:cf_subspace}
\end{equation}
discarding all bunched configurations; this models threshold detectors and is standard in boson sampling~\cite{aaronson_computational_2011,Zhong2020}. In the \emph{$r$-rail code} setting, $m=rN$ modes encode $N$ logical units with one photon per $r$-mode block; we study $r=2$ (dual-rail qubits~\cite{Kok2007,Knill2001}).

\subsubsection{Loss Function}
We adopt the Bhattacharyya loss~\cite{Bhattacharyya1943,Fuchs1999} between the postselected output distribution $p(g)$ and a target distribution $q$:
\begin{equation}
\mathcal{L}_{\mathrm{B}}(g; q) = 1 - \left(\sum_{x} \sqrt{p_x(g)\, q_x}\right)^2
\label{eq:bhat_loss}
\end{equation}

With $\{P_x\}_x$ the postselected measurement projectors and $O_x^{(K)}:=K^\dagger P_x K$, $O_{\mathrm{succ}}^{(K)}:=K^\dagger K$ the pulled-back observables, $p_x(g)=a_x(g)/s(g)$ with $a_x(g)=\mathrm{Tr}[\pi(g)\rho\pi(g)^\dagger O_x^{(K)}]$ and $s(g)=\mathrm{Tr}[\pi(g)\rho\pi(g)^\dagger O_{\mathrm{succ}}^{(K)}]$ — each an expectation value of the form Eq.~\eqref{eq:postselected_obj}, so the Lie-algebraic variance formula applies directly to these building blocks even though $\mathcal{L}_{\mathrm{B}}$ itself is nonlinear in the output distribution. Differentiating and eliminating $\partial_k p_x(g)$ via the quotient rule collapses the loss gradient into a single pulled-back observable,
\begin{equation}
  \partial_\mu\mathcal{L}_{\mathrm{B}}
  = \mathrm{Tr}\!\left[\partial_\mu\!\left(\pi(g)\rho\pi(g)^\dagger\right)
      O_{\mathrm{eff}}(\theta)\right],
  \label{eq:O_eff}
\end{equation}
with $O_{\mathrm{eff}} = \frac{1}{s(g)}\sum_x \kappa_x \left(O_x^{(K)} - p_x\,O_{\mathrm{succ}}^{(K)}\right)$, $\kappa_x = -B(p,q)\sqrt{q_x/p_x}$, and $B(p,q):=\sum_x\sqrt{p_xq_x}$ the Bhattacharyya coefficient. At each fixed parameter point the Bhattacharyya gradient therefore takes the form of a linear expectation-value gradient with effective observable $O_{\mathrm{eff}}(\theta)$. Because $O_{\mathrm{eff}}$ depends on the sampled circuit through $p_x(g)$, $s(g)$ and $B(p,q)$, it is correlated with $U$, and Eq.~\eqref{eq:photonic_var} does not directly give the unconditional Haar variance of the nonlinear loss gradient. We therefore use the irreducible support of $O_{\mathrm{eff}}$ as a local mechanistic diagnostic, and test the resulting scaling predictions independently by exact statevector simulation (full derivation in Appendix~\ref{app:bhat_local}).

\subsubsection{Initialization, Gradient Statistics, and Model Discrimination}
\label{sec:model_discrimination}

We compare three parameter initialization strategies to probe whether observed gradient scaling is an artifact of a particular initialization or a structural consequence of the postselection regime: \emph{uniform-random} (all mesh parameters sampled from $\mathrm{U}[0,2\pi)$), \emph{beta-random} (concentrating parameters near $0$ and $\pi$), and \emph{Haar-random} ($S\sim\mathrm{Haar}(U(m))$). For each configuration we draw $S=3000$ independent random circuits and targets. Targets are sampled as postselected outputs of random passive circuits so that they are reachable by the circuit family (Appendix~\ref{app:bhat_local}). Probabilities are regularized as $p,q \mapsto (1-\lambda)p + \lambda\,\mathrm{Unif}$ with $\lambda=10^{-3}$. We report the per-component empirical variance averaged over parameters,
\begin{equation}
\widehat{V}(N) = \frac{1}{P}\sum_{\mu=1}^{P}\mathrm{Var}_s[g^{(s)}_\mu]
\label{eq:var_est}
\end{equation}
computed via automatic differentiation through exact statevector simulation. 

To distinguish exponential concentration $\widehat{V}(N)\approx C\exp(-aN)$ from polynomial decay $\widehat{V}(N)\in O(N^{-c})$ from finite, noisy data, we apply four complementary tests uniformly to every configuration:

\begin{itemize}
\item \emph{Global fit and AICc:} both models fit by least squares on the common $N$-range; $\Delta\mathrm{AICc} = \mathrm{AICc}_{\mathrm{poly}} - \mathrm{AICc}_{\mathrm{exp}}$, with $\Delta\mathrm{AICc}<0$ favoring the polynomial model.
\item \emph{Sliding-window exponents:} local $c(N)$ and $a(N)$ fit separately over overlapping $8$-point windows; whenever the polynomial model is favored we further check $n$ apparent local exponential rate is an artifact of $c/N$ rather than an actual rate.
to within a few percent on late windows.
\item \emph{Truncation sweep:} the global fit is repeated after discarding the smallest points, $\mathrm{min\_qubits}\in\{1,2,3\}$, testing stability of the preferred model and fitted exponents.

\item \emph{Crossover model and status labels:} fitting the two-regime model $\widehat{V}(N) = C\,N^{-c}\,e^{-aN}$ extracts the crossover scale $N^{*}$ at which the polynomial and exponential contributions become comparable (\texttt{interpolated} if $N^*$ within the simulated range, \texttt{extrapolated\_below}/\texttt{\_above} if $N^*$  below or above it, or \texttt{unidentified} otherwise).

\end{itemize}

Given the cost of the $S=3000$-sample estimator repeated across nine configurations, we do not report per-window bootstrap confidence intervals. The agreement among the four tests above is the primary evidence for each verdict, with bootstrap-based interval estimation left to future work. 

Finally, we validated the estimation pipeline itself: we reproduced the standard gate-based BP on random parameterized circuits of $5$-$18$ qubits at depths $20$-$120$, matching results in ~\cite{McClean2018}. The MerLin framework was used for the photonic experiments~\cite{notton2026merlin} and the code is available on \href{https://github.com/easonoob/Eason-2026-preasymptotic-trainability-pvqc}{GitHub}.

\section{Experiments}

\subsection{Postselection Regime Comparison}
\label{sec:regime_comparison}

We evaluate gradient statistics for the nine configurations (3 postselection regimes each under the 3 initialization strategies. Due to simulation cost, the accessible range differs by regime: Fock is evaluated over $N = 2$-$10$, Unbunched over $N = 2$-$13$; Dual-rail, whose postselected subspace has dimension $2^N$, scales to $N = 24$. This asymmetry in accessible range is a limitation for the Fock and Unbunched verdicts and is discussed further in Section~\ref{sec:model_discrimination} and in the limitations. We focus on per-parameter variance rather than aggregate norm, as the norm can be confounded by parameter-count growth under code postselection.

\begin{figure*}[ht]
\centering
\includegraphics[width=0.9\linewidth]{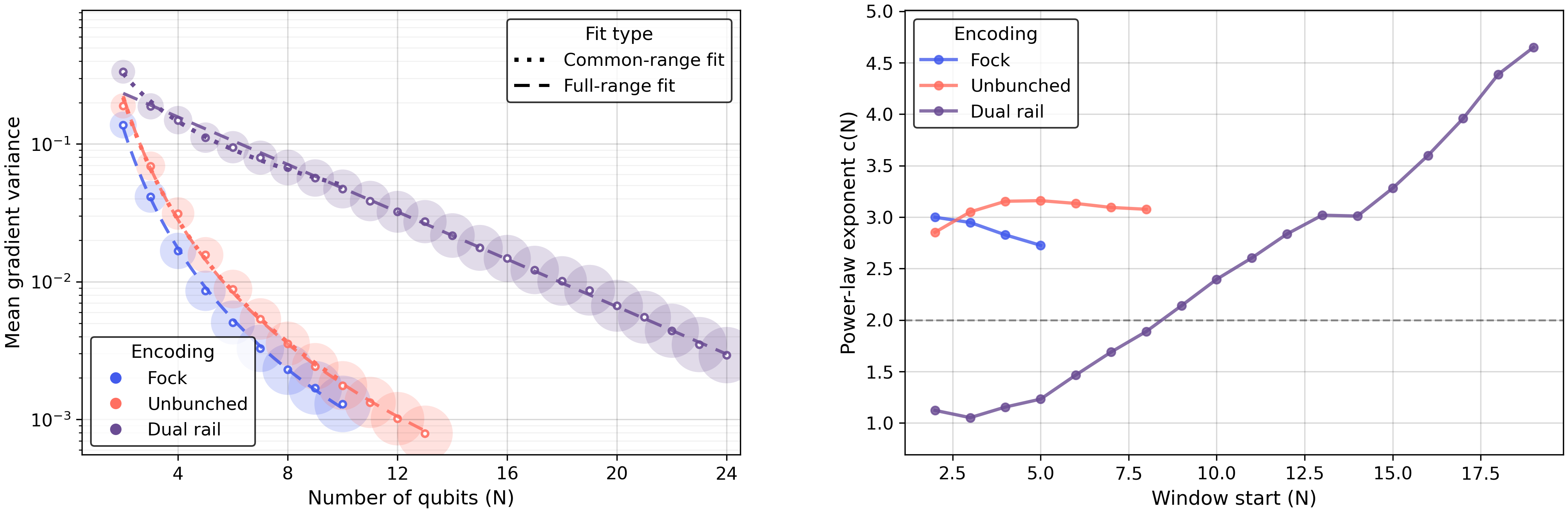}
\caption{Gradient-variance scaling for Haar-random initialization. \emph{Left:} mean gradient variance $\widehat{V}(N)$ versus $N$ for the three postselection regimes; circle area is proportional to postselected subspace dimension. Dotted lines show the common-range fit ($N=2$-$10$) used for cross-regime comparison; dashed lines show the full-range fit over each regime's accessible sizes. \emph{Right:} sliding-window power-law exponent $c(N)$ (8-point windows) for each regime; the dashed reference line marks $c=2$. A flat $c(N)$ reliably signals true polynomial decay, whereas a growing $c(N)$ indicates underlying exponential decay.} 
\label{fig:main_result}
\end{figure*}

We report results for Haar-random initialization in Figure~\ref{fig:main_result}, while Table~\ref{tab:verdict_summary} provides the model-discrimination outcome for all 9 configurations (Section~\ref{sec:model_discrimination}). In light of the bridge developed in Section~\ref{sec:math-fram}, we interpret these experiments as testing whether the observable-support mechanism identified by the Lie-algebraic framework persists for the nonlinear Bhattacharyya loss, rather than as a direct theorem-to-data comparison for Eq.~\eqref{eq:photonic_var}.

\textbf{Fock.} For circuits with $N\leq10$, we find no evidence of exponential concentration, regardless of initialization or truncation. From Table~\ref{tab:verdict_summary}, we conclude that the polynomial model is preferred by $\mathrm{AICc}$. This is corroborated by the sliding-window rate $a(N)$ which decays monotonically rather than stabilizing (e.g.\ $0.603\to0.477$, Haar). No initialization yields a crossover scale $N^{*}$ within the simulated range at default truncation, consistent with the absence of exponential concentration over the tested range. This result should not be interpreted as evidence of trainability at larger $N$.

\textbf{Unbunched.} Similarly to the Fock regime, the data are polynomial-consistent (Table~\ref{tab:verdict_summary}), with reproducible small-$N$ curvature that fades as $N$ grows. Over sliding $8$-point windows, $c(N)$ plateaus rather than growing without bound (e.g.\ $2.93\to3.14\to3.11$, Haar). The identity $a(N)\,N \approx c$ holds to a few percent whenever the polynomial model is favored (Haar: $a\bar N=3.12$ vs.\ fitted $c=3.08$), confirming the apparent exponential rate is in fact $c/N$. The crossover $N^{*}$ instead diverges as small points are dropped (Haar: $13.1\to41.8\to156.2$): this would be the fitting artifact of the transient. We note, however, that polynomial decay with $c \approx 3$ does not imply unconditional trainability: the associated shot cost scales as $N^{3}$, so training in this regime remains polynomially expensive.

\textbf{Dual-rail.} From Table~\ref{tab:verdict_summary} and Figure~\ref{fig:main_result}, we observe that the three initializations yield nearly identical curves and results, indicating that concentration is a structural consequence of rail-code postselection. The polynomial model is nominally favored on the common
range $N=2$-$10$, but the verdict flips decisively to exponential on the full range $N=2$-$24$: the short-range window cannot distinguish the two models, and only the extended range resolves the trend. The full-range rate, $a \approx 0.194$-$0.198$ nats/qubit (one decade per $\approx 11.6$-$11.9$ qubits), is stable to within $0.17$-$0.24$ across $15$-$16$ overlapping windows spanning $N=2$-$24$, whereas the local power-law exponent $c(N)$ grows without bound. Consistently, $\Delta\mathrm{AICc}$ \emph{grows} with truncation (Haar: $+42.9 \to +64.1 \to +94.8$), opposite the Unbunched trend, and the fitted $N^{*}$ is \texttt{extrapolated\_below} or negative in $8$ of $9$ cells. We conclude that concentration is already present at our smallest simulated size.

Under the operational criterion of \cite{McClean2018}, dual-rail exhibits a BP, but at a rate roughly an order of magnitude below the $2$-design value: $a \approx 0.19$--$0.20$ nats per logical qubit, versus $2\ln 2$--$3\ln 2 \approx 1.39$--$2.08$ for Haar-random circuits at an approximate $2$-design. Moreover, the circuits are depth one, with $\mathfrak{u}(m)$ of dimension $m^2 = 4N^2$. The concentration therefore arises from postselection projecting the effective observable into high-dimensional operator modules (Section~\ref{sec:math-fram}). At $N=24$, variance remains $\widehat V \sim 10^{-2}$--$10^{-3}$, against $\sim 10^{-14}$ at the $2$-design rate. This is exponential in the asymptotic sense, but far from operationally severe.

\begin{table}[t]
\centering
\resizebox{\linewidth}{!}{
\begin{tabular}{llccccl}
\toprule
\textbf{Regime} & \textbf{Init} & $N_{\max}$ & $c$ & $a$ & $\Delta$AICc & $N^{*}$ \textbf{status} \\
\midrule
Fock & Haar & 10 & 2.91 & 0.55 & $-37.8$ & unidentified \\
Fock & Beta & 10 & 2.90 & 0.54 & $-23.5$ & unidentified \\
Fock & Unif. & 10 & 2.77 & 0.53 & $-52.4$ & unidentified \\
Unbunched & Haar & 13 & 2.99 & 0.47 & $-40.4$ & extrap.\ above ($41.8$) \\
Unbunched & Beta & 13 & 3.30 & 0.52 & $-31.2$ & extrap.\ above ($231.0$) \\
Unbunched & Unif. & 13 & 2.61 & 0.41 & $-35.2$ & extrap.\ above ($35.7$) \\
Dual-rail & Haar & 24 & 1.88 & 0.198 & $+64.1$ & extrap.\ below ($1.38$) \\
Dual-rail & Beta & 24 & 1.80 & 0.196 & $+51.9$ & extrap.\ below ($1.37$) \\
Dual-rail & Unif. & 24 & 1.82 & 0.194 & $+84.3$ & extrap.\ below ($0.31$) \\
\bottomrule
\end{tabular}%
}
\caption{Model-discrimination summary at default truncation ($\mathrm{min\_qubits}=2$). $c$ and $a$ are the fitted power-law exponent and exponential rate (nats) on each regime's full accessible range; $\Delta\mathrm{AICc} = \mathrm{AICc}_{\mathrm{poly}}-\mathrm{AICc}_{\mathrm{exp}}$ (negative favors polynomial); $N^{*}$ status is defined in Section~\ref{sec:model_discrimination}; for all runs, $m=2N, n=N$.}
\label{tab:verdict_summary}
\end{table}
\subsection{Mode Count Scaling under Collision-Free Postselection}
\label{sec:modes_sweep}

To characterize whether the polynomial regime depends on the mode count beyond the minimum encoding $m = N$, we additionally sweep the total mode count $m$ from $2N$ to $24$ under Unbunched postselection, for fixed $N \in \{5,\ldots,12\}$. Gradient variance is only mildly separated across $N$ at fixed $m$ and does not diverge exponentially, and curves for adjacent $N$ converge at large $m$; the polynomial character of this regime is therefore robust to ancillary mode count, consistent with Prediction II.1.

\section{Discussion}
\label{sec:discussions}

We show a distinction between two finite-size trainability regimes in passive photonic variational circuits (Table~\ref{tab:verdict_summary}). The polynomial verdicts in the allow-bunching and collision-free settings align with the predictions of Section~\ref{sec:math-fram}: passive linear optics explores a $U(m)$ representation whose relevant algebraic dimension grows only as $m^2-1$, so increasing system size does not by itself force the strong concentration of standard BP. We therefore read the data as finite-size evidence for mild variance dilution~\cite{Ragone2024,Larocca_2025,Meyer2023,Monbroussou_2025}. 

The dual-rail verdict points to a different mechanism. Code postselection changes the pulled-back observables $K^\dagger P_x K$ and $K^\dagger K$: dual-rail postselection selects a subspace isomorphic to $(\mathbb{C}^2)^{\otimes N}$, concentrating $(O^{(K)})$ into high-index irreducible modules of exponentially growing dimension. This is consistent with the broader picture that concentration depends on the joint structure of circuit family, input state, and observable~\cite{McClean2018,Larocca_2025,Holmes2022}. Measured against the operational criterion of ~\cite{McClean2018}, this places dual-rail circuits in the BP regime, but at a rate roughly an order of magnitude below the $2$-design value.

An additional observation qualifies this contrast. Under a substantially more mixing ensemble, collision-free postselection also exhibits exponential-like gradient decay. Because those experimental settings differ from the principal comparison, this result does not provide a controlled comparison of crossover thresholds or decay rates. It nevertheless indicates that the observed trainability scaling depends jointly on postselection geometry and ensemble mixing, rather than supporting a general universality–trainability trade-off.

Within the principal matched ensembles, the contrast between collision-free filtering and code postselection remains informative: a smaller postselected space alone is insufficient to trigger a BP. The results instead suggest that gradient concentration depends on how the postselection map reshapes the effective observables together with the degree of ensemble mixing.

These results carry several limitations. Conclusions rest on finite-size exact simulations, not asymptotic proofs as $O^{(K)}$ offers a mechanistic reading of the nonlinear Bhattacharyya loss rather than a closed-form theorem. In particular, the exact variance formula applies to linear expectation-value costs with fixed observables; its use for the Bhattacharyya loss is mechanistic rather than a closed-form result, since $O_{eff}(\theta)$ is itself circuit-dependent. Simulations are noiseless and ignore postselection success probability, so a resolvable gradient may still need exponentially more trials as system size grows, and other noise sources could independently induce concentration~\cite{Wang2021NIBP,Maring2024,Baldazzi2025}. The BP language is not a classical-hardness claim: neither additive permanent approximations~\cite{gurvits2005complexity} nor proven boson-sampling hardness~\cite{aaronson_computational_2011} (valid only for $m=\omega(n^2)$, not our $m=2N$) apply here. We also omit confidence intervals given the cost of resampling across nine configurations, relying instead on the four-test agreement of Section~\ref{sec:model_discrimination} as corroboration. Finally, the Fock ceiling ($N\leq10$) weakens its negative result, and as depth is not an independently tunable axis here, our polynomial verdicts should not be read as depth-independent for other, non-universal photonic architectures where depth-driven expressivity could still induce concentration.

\section{Conclusion}

We studied trainability in passive photonic variational circuits via exact statevector simulation across postselection regimes. Gradient variance decays polynomially under allow-bunching and collision-free filtering but turns exponential under dual-rail postselection beyond moderate system sizes. We showed that this result comes from how each map reshapes the pulled-back observable across irreducible modules. Since MZI depth does not enlarge the reachable $U(m)$ family, this result reflects postselection geometry.

Favorable scaling elsewhere is not yet a hardware guarantee: photon loss, detector efficiency, distinguishability, and finite-shot cost remain open factors for a complete assessment. We leave these limitations for future works.

\section*{Acknowledgement}
The authors thank R. Mezher, B. Ventura,  P-E. Emeriau, H. Thomas, L. Monbroussou and H. Mhiri for helpful discussions and insightful feedback during the development of this project, and acknowledge the use of AI-based tools for language polishing and code debugging assistance. All scientific content and results were verified by the authors.

\appendices
\section{Computation of the Irreducible Projections $P_k^{(n)}$}
\label{app:gpurity_computation}

The raising/lowering maps $R,L$ of Section~\ref{sec:math-fram} give the recursive projector onto $\lambda_k^{(n)}$,
\begin{equation}
\small
  P_k^{(n)}(O) = \frac{1}{\alpha_{kk+1,n}^k}
    \left(R^{n-k}\!\circ L^{n-k}(O) - \sum_{r=0}^{k-1}\alpha_{k+1,n}^r\,P_r^{(n)}(O)\right),
  \label{eq:irrep_projection}
\end{equation}
where $\alpha_{k+1,n}^r = \frac{(n-r)!\,(n+m+r-1)!}{(k-r)!\,(k+m+r-1)!}$, for $O \in W_n$. 

We implement $R,L$ on the Fock sectors $\mathcal{F}_p$ enumerated with the basis/dimension utilities of QOptCraft~\cite{aguado2023qoptcraft}. We follow Eq.~ (15-16) in \cite{mhiri2026boson} with $A_i^{(p)}$ the matrix of the mode-$i$ annihilation operator $a_i\ket{s}$.

Algorithm~\ref{algo:procedure_variance} builds $\{A_i^{(p)}\}$ once, applies $R^{n-k}\circ L^{n-k}$ (Eq.~\eqref{eq:irrep_projection}) to project $\rho$ and $O^{(K)}$ onto all $n+1$ modules, and evaluates Eq.~\eqref{eq:photonic_var}.

\begin{algorithm}[ht]
\caption{Exact Gradient Variance $\mathrm{Var}_U[f]$ via Irreducible Projections}
\label{algo:procedure_variance}
\begin{algorithmic}[1]
\small
\Require Modes $m$, photons $n$, observable $O$, input state $\rho$, post-selected indices $\mathcal{I}$
\Ensure $\mathrm{Var}_U[f]$ of Eq.~\eqref{eq:photonic_var}

\State Build $A_i^{(p)}:\mathcal{F}_p\to\mathcal{F}_{p-1}$, $i=1,\dots,m$, $p=1,\dots,n$
\State $O^{(K)} \gets \Pi_\mathcal{I}\, O\, \Pi_\mathcal{I}$ \Comment{embed restricted block $O[\mathcal{I},\mathcal{I}]$, zero outside $\mathcal{I}$}

\Function{Project}{$X \in W_n$}
    \Comment{returns $\{P_k^{(n)}(X)\}_{k=0}^n$, Eq.~\eqref{eq:irrep_projection}}
    \For{$k = 0$ to $n$}
        \State $Y \gets X$
        \For{$p = n$ downto $k+1$}
            \State $Y \gets \sum_i A_i^{(p)}\,Y\,(A_i^{(p)})^\dagger$
        \EndFor
        \For{$p = k$ to $n-1$}
            \State $Y \gets \sum_i (A_i^{(p+1)})^\dagger\,Y\,A_i^{(p+1)}$
        \EndFor
        \State $P_k \gets \dfrac{1}{\alpha(k,k)}
        \left(
        Y - \sum_{r=0}^{k-1} \alpha(r,k)\,P_r
        \right)$
    \EndFor
    \State \Return $\{P_0,\dots,P_n\}$
    \Comment{$\alpha(r,k) := \dfrac{(n-r)!\,(n+m+r-1)!}
    {(k-r)!\,(k+m+r-1)!}$}
\EndFunction

\State $\{P_k(\rho)\}_{k=0}^n \gets \Call{Project}{\rho}$
\State $\{P_k(O^{(K)})\}_{k=0}^n \gets \Call{Project}{O^{(K)}}$

\For{$k = 0$ to $n$}
    \State $w_k(\rho) \gets \|P_k(\rho)\|_2^2$
    \State $w_k(O^{(K)}) \gets \|P_k(O^{(K)})\|_2^2$
    \Comment{$d_k^{(n)}$ from Eq.~\eqref{eq:irrep_decomp}}
\EndFor

\State \Return $\mathrm{Var}_U[f] \gets
\displaystyle\sum_{k=1}^{n}
\dfrac{w_k(\rho)\,w_k(O^{(K)})}{d_k^{(n)}}$

\end{algorithmic}
\end{algorithm}

Only $k\in\{0,1\}$ occurs for $n=1$, recovering the single-module $\mathfrak{su}(m)$-purity formula; for $n>1$ the modules $k\ge2$ lie outside the DLA image $\pi_{m,n}(\mathfrak{u}(m))$ and generally carry non-negligible weight, so all $n+1$ terms are needed. The implementation is validated against the sum rule $\sum_{k=0}^n w_k(H) = \|H\|_2^2$, $H=\rho,O^{(K)}$.

\section{Local Expansion of the Loss and Link to $\mathrm{Var}_U[f]$}
\label{app:bhat_local}

The variance formula in ~\ref{sec:math-fram} is exact for linear expectation-value costs, but the experiments minimize the nonlinear Bhattacharyya loss $\mathcal{L}_{\mathrm{B}}$. Here we (i) derive the exact loss gradient underlying Eq.~\eqref{eq:O_eff}, and (ii) show that $\mathcal{L}_{\mathrm{B}}$ reduces locally to a linear-cost problem at fixed $\theta$, isolating the pulled-back observable $O_{\mathrm{eff}}$ used to compute the gradient statistics.

With $p_x(g)=a_x(g)/s(g)$ as in \ref{sec:math-fram}, the quotient rule gives
\begin{equation}
\partial_k p_x(g)
=\frac{1}{s(g)}
\left(\partial_k a_x(g)-p_x(g)\,\partial_k s(g)\right).
\label{eq:px_grad}
\end{equation}
For outcomes with $p_x(g)>0$, the loss gradient is exactly
\begin{equation}
\partial_k \mathcal{L}_{\mathrm{B}}(g;q)
=-B(p(g),q)\sum_x \sqrt{\frac{q_x}{p_x(g)}}\,\partial_k p_x(g),
\label{eq:bhat_grad}
\end{equation}
an exact parameter-dependent linear combination of gradients of expectation values of $O_x^{(K)}$ and $O_{\mathrm{succ}}^{(K)}$. Substituting~\eqref{eq:px_grad} into ~\eqref{eq:bhat_grad}, collecting terms in $\partial_k a_x(g)$ and $\partial_k s(g)$ gives Eq.~\eqref{eq:O_eff}.

Near a reachable reference $q$, write $p_x=q_x+\delta p_x$ with $\sum_x\delta p_x=0$, $|\delta p_x|\ll q_x$. Expanding $\sqrt{p_xq_x}$ to second order and using $\sum_xq_x=1$,
\begin{equation}
\mathcal{L}_{\mathrm{B}}(p,q) = 1-B(p,q)^2 \approx \tfrac14\sum_x\frac{\delta p_x^2}{q_x} = \tfrac14\,\chi^2(p\|q),
\end{equation}
up to $O(\|\delta p\|^3)$: locally, $\mathcal{L}_{\mathrm{B}}$ is a $\chi^2$-divergence. The hypothesis $|\delta p_x|\ll q_x$ requires $q$ to be reachable by the circuit family: a target drawn uniformly on the postselected simplex makes $q_x/p_x$ heavy-tailed and breaks the expansion. This is why targets are sampled as postselected circuit outputs with regularized probabilities, as described in Section~\ref{sec:model_discrimination}.

The same expansion applied to $\kappa_x=-B(p,q)\sqrt{q_x/p_x}$ in Eq.~\eqref{eq:O_eff} gives $\kappa_x=-1+\delta p_x/2q_x+O(\delta^2)$. Since $\sum_xO_x^{(K)}=O_{\mathrm{succ}}^{(K)}$, the constant term cancels, leaving
\begin{equation}
O_{\mathrm{eff}} = \frac{1}{2s(g)}\sum_x\frac{\delta p_x}{q_x}\Big(O_x^{(K)}-p_x\,O_{\mathrm{succ}}^{(K)}\Big)+O(\delta^2),
\label{eq:O_eff_local}
\end{equation}
which reduces the Bhattacharyya gradient to a linear-cost problem at fixed $\theta$ and is used directly to compute the gradient statistics reported in the main text.

\newpage
\bibliographystyle{ieeetr}
\bibliography{references}

\end{document}